\documentstyle[12pt]{article}
\def\bold#1{\setbox0=\hbox{$#1$}%
      \kern-.025em\copy0\kern-\wd0
      \kern.05em\copy0\kern-\wd0
      \kern-.025em\raise.0433em\box0 }
\def\nsp{\noindent}
\def\eea{\end{eqnarray}}
\def\bea{\begin{eqnarray}}
\def\eeas{\end{eqnarray*}}
\def\beas{\begin{eqnarray*}}
\def\ee{\end{equation}}
\def\be{\begin{equation}}
\def\bdm{\begin{displaymath}}
\def\edm{\end{displaymath}}
\def\fr{\frac}

\def\rd{{\rm d}}

\def\veps{\varepsilon}

\def\dag{^\dagger}

\def\fpi2{\mbox{F$_\pi$}^2}

\def\mpi2{{m_\pi}^2}
\def\mk{m_K}
\def\mk2{{m_K}^2}

\def\fk2{\mbox{F$_K$}^2}

\def\Tr{{\,\mbox{Tr}}}

\begin{document}
\begin{titlepage}
\begin{center}
\vspace*{2.0cm}
\hfill hep-ph/9309305
{\large\bf BOUND STATE SOLITON DESCRIPTION OF LOW PARTIAL WAVE
OCTET BARYON RESONANCES}
\vskip 1.5cm
{\large CARLO GOBBI}
\vskip .2cm
{\large \it Center for Theoretical Physics, Sloane Physics Laboratory,}\\
{\large \it   Yale University, New Haven, Connecticut 06511, USA}\\
{\large \it  and INFN, Sezione di Pavia, Pavia, Italy}
\vskip .2cm
and
\vskip .2cm
{\large NORBERTO N. SCOCCOLA}
\vskip .2cm
{\large \it
Physics Department, Comisi\'on Nacional de Energ\'{\i}a At\'omica,}\\
{\large \it Av.Libertador 8250, (1429) Buenos Aires, Argentina}
\vskip 2.cm
{\bf ABSTRACT}\\
\begin{quotation}
A version of the bound state soliton model which allows both $\eta$ and $K$
bound states is used to study low partial wave octet baryon resonances.
It is found that negative parity $S$--wave resonances are well described
within this framework. A possible interpretation of the $P$--wave
resonances is also discussed.
\end{quotation}
\end{center}
\end{titlepage}

Although the idea of describing some of the nucleon resonances as
eta-nucleon bound state was proposed long ago \cite{GT65}, only
recently this possibility was investigated in the context of the soliton
models. In Ref.\cite{PAR91} it was shown that when an alternative
fourth order term is included together with the usual Skyrme term,
an $S$--wave eta-soliton bound state appears.
This eta-soliton bound state has been identified with the
N(1535) $S$--wave resonance,  which has a large branching
ratio for decays into $\eta N$.
The introduction of the
alternative fourth order term is justified
in the framework of chiral perturbation theory \cite{GL85}. Moreover,
some constraints on the strength of such a term with respect to the one
of the Skyrme term can be obtained from phenomenological sources
\cite{PS92}.
The validity of this scheme has been tested by the
calculation of the eta photoproduction amplitudes from nucleons
\cite{GHS92}.  The purpose of
the present article is to extend this picture to the description of those
strange baryon resonances which have a relative large branching ratio for
the decay into the corresponding low lying
hyperon state and an $\eta$.  As in Ref.\cite{PAR91} our work
is based on the bound state soliton approach proposed by Callan and
Klebanov \cite{CK85}, which has been shown to be very successful in the
description of hyperon properties \cite{RS91,OMRS91}.

We start with the effective soliton Lagrangian with an
appropriate symmetry breaking term, expressed in terms
of the $SU(3)$--valued chiral field $U(x)$ as
\bea
\Gamma = \Gamma^{(2)}+\Gamma^{(4)} + \Gamma_{WZ} + \Gamma_{SB}.  \label{action}
\eea
\nsp
$\Gamma^{(2)}$ is the usual kinetic term
\bea \label{2}
\Gamma^{(2)} & =
   & - {f_\pi^2\over4}
   \int \rd^4x \ \Tr (L_\mu L^\mu)
\eea
while $\Gamma^{(4)}$ is the fourth
order interaction term written as
\bea \label{4}
\Gamma^{(4)} & =
   & \int \rd^4x \left\{ {x \over {32e^2}} \Tr [L_\mu, L_\nu]^2
   +    {{1-x} \over {16e^2}} \{ (\Tr L_\mu L_\nu)^2
                                -(\Tr L_\mu L^\mu)^2 \} \right\} ,
\eea
where $L_\mu=U^\dagger \partial_\mu U$, $U$ being the chiral soliton
field.

The first term in RHS of Eq.(\ref{4}) is the standard quartic Skyrme
term  ($e$ is the so--called Skyrme parameter), that yields no interaction
between the soliton and the $\eta$ field.
The second contribution is the ``alternative" quartic term introduced
in Ref.\cite{PAR91}. The parameter $x$ weights the relative strength
of these two terms, its range going from 0 to 1.

In Eq.(\ref{action}) $\Gamma_{SB}$ is responsible for the explicit
breaking of chiral symmetry. We use the following form for
$\Gamma_{SB}$
\bea
\Gamma_{SB}
& \! \! = \! \! &\int \rd^4x \left\{ {f_\pi^2  m_\pi^2 + 2 f_K^2 m_K^2
\over{12}}
\Tr \left[ U + U^\dagger - 2 \right] \right. \nonumber \\
& & \hskip .5cm + { f_\pi^2 m_\pi^2 - f_K^2 m_K^2 \over{6}}
\Tr \left[ \sqrt{3} \lambda^8 \left( U + U^\dagger \right) \right]
\nonumber \\
& & \hskip .5cm \left. - { f_K^2 - f_\pi^2 \over{12} } \Tr
\left[ \left( 1 - \sqrt{3} \lambda^8 \right)
\left( U \partial_\mu U^\dagger \partial^\mu U +
        U^\dagger \partial_\mu U \partial^\mu U^\dagger \right) \right]
\right\} ,
\label{oldmass}
\eea
\noindent
where $\lambda^8$ is the eighth Gell-Mann matrix,
$f_\pi$ is the pion decay constant (~$= 93$ MeV
empirically), $f_K$ is the kaon decay constant
and $m_\pi$ and $m_K$ represent the pion and kaon masses respectively.
Eq.(\ref{oldmass}) accounts for both the finite mass of the pseudoscalar mesons
and the empirical difference between their decay constants.
Finally, $\Gamma_{WZ}$ is the non--local Wess--Zumino action
\be
\Gamma_{WZ} \ = \ -i \fr{N_c}{240\pi^2}\int \ \rd^5x \
\veps^{\mu\nu\alpha\beta\gamma}
\  \Tr (L_\mu L_\nu L_\alpha L_\beta L_\gamma) \, .
\ee

In order to describe the eta-kaon-soliton system we introduce a generalized
Callan-Klebanov ansatz for the chiral field given by
\be
\label{ansatz}
U=\sqrt{U_\pi} U_{K} U_{\eta} \sqrt{U_\pi} \ ,
\ee
where
\bea
U_K \ = \ \exp \left[ i\fr{\sqrt2}{{f_K}} \left( \begin{array}{cc}
                                                        0 & K \\
                                                        K\dag & 0
                                                   \end{array}
                                           \right) \right] \ , \
                                           \ \ \
K \ = \ \left( \begin{array}{c}
                   K^+ \\
                   K^0
                \end{array}
                           \right),
\eea
and
\bea
U_{\eta} \ = \ \exp \left[ i {\eta \lambda_8 \over{f_\eta}} \right]\, .
\eea
Finally, $U_\pi$ is the soliton background field written as
a direct extension to $SU(3)$ of the $SU(2)$ field $u_\pi$, i.e.,
\bea
U_\pi \ = \ \left ( \begin{array}{cc}
                       u_\pi & 0 \\
                       0 & 1
                    \end{array}
                               \right ) \ .
\eea

It should be noticed that the $\eta$ field has been introduced as the
eight component of the pseudoscalar octet. A similar approximation
has been used in previous calculation \cite{PAR91,GHS92}.
Considering the rather small mixing angle in the physical $\eta$ this
appears as a reasonable approximation.

As mentioned above, our effective action includes chiral symmetry breaking
effects that lead to different decay constants for the different
pseudoscalar mesons. In particular, $f_\eta$ is related to the
pion and kaon decay constants by
\be
f_\eta^2 = {4\over3} f_K^2 - {1\over3} f_\pi^2 \, .
\ee
Using the empirical ratio $\chi_K = f_K/f_\pi = 1.22$,
one gets $f_\eta/f_\pi = 1.28$.
This value agrees well with the estimation given in Ref.\cite{GK87}.
Within the present model
the $\eta$ mass is given in terms of the pion and kaon
masses by
\bea
m_\eta^2 = { 4 \chi_K^2 m_K^2 - m_\pi^2 \over{ 4 \chi_K^2 - 1 } }\, .
\eea
One can easily see that for $\chi_K \ne 1$ there are corrections
to Gell-Mann--Okubo mass formula. Using the empirical values of $\chi_K$,
$m_\pi$ and $m_K$ we obtain $m_\eta = 539$ MeV to be compared
with the empirical value $m_\eta^{emp} = 549$ MeV

Following the usual steps of the bound state approach we expand
up to second order in the kaon and eta field. To this order, these
fields are decoupled one from the other, their interactions being only
with the soliton background
field. Therefore, the kaon-soliton effective action
reduces to the one discussed in Ref.\cite{PS92}. One the other hand,
the resulting eta-soliton effective action has some differences with
respect to the one used in Ref.\cite{PAR91}. This is due to the more
realistic symmetry breaking action used in the present calculation.
Using the hedgehog ansatz for the $SU(2)$ soliton field
\be
u_\pi = \exp\left[ i \ {\bold \tau} \cdot \hat {\bold r} \ F(r) \right]
\ee
our eta-soliton action to ${\cal O}(N_c^0)$ reads
\bea
L_{\eta-sol} = {1\over2} \int \rd^3x \ \left\{
\alpha \dot \eta^2 - \left[ \rho ( {\bold \nabla} \eta )^2
- (\beta - \rho) ( \partial_r \eta )^2 \right]
+ \gamma^2 \eta^2 \right\} \, ,
\eea
where
\bea
\alpha &=& 1 + {1-x\over{e^2 f_\eta^2} } \left[ F'^2 +
2 {\sin^2 F \over{r^2}} \right]\, , \\
\beta &=& 1 +  {1-x\over{e^2 f_\eta^2} } \, 2 {\sin^2 F \over{r^2}}\, , \\
\rho &=& 1 +  {1-x\over{e^2 f_\eta^2} }
\left( F'^2 + {\sin^2 F \over{r^2}} \right) \, , \\
\gamma^2 &=& m_\eta^2 - { m_\pi^2 f_\pi^2 \over{3 f_\eta^2} } \ (1- \cos F)
\, .
\eea
Using the standard partial-wave decomposition of the $\eta$ field
we find that the eingenmodes that diagonalize the hamiltonian
in each partial wave can be written as
\bea
\eta ({\bold r},t) = e^{-i \omega_\eta t} \ \eta(r)\, Y_{lm}(\theta,\phi) \, ,
\eea
where $\eta(r)$ satisfies the eigenvalue equation
\bea
\left\{ {1\over{r^2}} \partial_r \left[ r^2 \beta \partial_r \right] +
\alpha \,\omega_\eta^2 - {l(l+1) \over{r^2}} \rho - \gamma^2
\right\} \eta = 0 .
\label{etaeq}
\eea
In our numerical
calculations we use two sets of values for the parameters
appearing in the effective lagrangian. In SET A we keep $f_\pi$, $f_K$
and $m_\pi$ fixed to their empirical values. SET B corresponds to the case
often used in skyrmion physics in which the pion is considered
as massless and $f_\pi$ takes the value $f_\pi = 64.5$ MeV.
In the latter case, the ratio $f_K/f_\pi$
is taken to its empirical value (= 1.22). In both cases the value of
$e$ is adjusted to reproduce the empirical $\Delta-N$ mass difference
and $m_K$ is taken to the empirical value $m_K = 495$ MeV.
As already mentioned the kaon sector of the present model
has been studied in Ref.\cite{PS92}. In particular, the kaon
binding energies $\omega_K$ and hyperfine splitting constants $c_K$ have
been calculated for different values of the mixing parameter $x$.
It was found that the hyperon masses are better reproduced when the
values $x=0.33$ (for SET A) and $x=0.66$ (for SET B) are used.

Using the parameter sets given above we have solved the $\eta$
eigenvalue equation Eq.(\ref{etaeq}). As in previous calculations
\cite{PAR91,GHS92} we have found that the only bound
state appears in the $S$--wave. The corresponding eigenenergy $\omega_\eta$
as a function of $x$ for both SET A and SET B is shown in
Fig.1. When comparing our results with those of Ref.\cite{PAR91}
we observe that the eta is less bound in our calculation. This
is mainly due to the incorporation of symmetry breaking terms that
take into account the difference between the $\eta$ decay constant
and the pion one. A similar effect was found in the kaon sector
\cite{RS91}.
Using the values of $x$ that lead to the best agreement with
empirical ground state hyperon masses we find $\omega_\eta (x=0.33)
=467$ MeV (SET A) and $\omega_\eta (x=0.66)
=517$ MeV (SET B).

In order to obtain the correct baryon quantum numbers we have to
consider the $SU(2)$ soliton isospin rotations. This leads to
the hyperfine corrections to the masses. Since the eta is bound
in an $S$--wave the corresponding hyperfine splitting constant
vanishes to ${\cal O}(N_c^{-1})$. Therefore, the mass formula
for spin 1/2 particles reads:
\bea
M  =  M_{sol} + n_K \omega_K + n_\eta \omega_\eta +
{1\over{2 {\cal I}}} \left[ (1-c_K) I (I+1) + {3\over4} c_K^2 \right]\, ,
\eea
where $M_{sol}$, $\cal I$ and $I$ are respectively the soliton mass,
moment of inertia and isospin. $\omega_K$ and $c_K$ are the kaon
energy and hyperfine constant, $n_K$ and $n_\eta$ represent the number of bound
kaon(s) and $\eta$ meson(s), related to the strangeness and parity quantum
numbers, respectively.

In Table 1 we list the calculated masses of the 1/2 baryon octet
together with the corresponding $S$--wave negative parity resonances.
Also shown are the empirical values taken from Ref.\cite{PDT}.
In the case of the $\Xi(1950)$ the assignment is only tentative since the
spin and parity of this state have not been empirically determined yet.
The values of $\omega_K$ and $c_K$ used to construct this table have
been taken from Ref.\cite{PS92}.

Apart from the negative parity $S$--wave resonances there are some
other low-partial wave baryon resonances which also have a rather
large $\eta$-baryon branching ratio.
An example of such states is the $P_{13}$
N(1710) resonance. To investigate whether these states
could be described within the present model we have
solved the $\eta$-eigenvalue equation, Eq.(\ref{etaeq}), in the
continuum for different partial waves.
The corresponding phase shifts as a function of the
eta energy $\omega_\eta$ are shown in Fig.2.
We plot the phase shifts for the lower partial waves for both sets of
parameters. As we see, the $l=0$ waves show the typical behaviour due
to the presence of a bound state. On the other hand,
%Fig.2-a corresponds
%to SET A ($x=0.33$) while Fig.2-b to SET B ($x=0.66$).
although the interaction
is attractive in all the other partial waves, there is no resonance even
for $l=1$. Similar results are obtained for other values of the
$x$-parameter. At this point it is important to recall that, within
the bound state model, most of the hyperon resonances with $l \ge 2$
can be understood as kaon-soliton resonance states \cite{SCO90}.

The results of Table 1 show that a comprehensive description of low--lying
$S$--wave baryon resonances
in terms of eta-kaon-soliton bound states is reliable.
The presence of a bound $\eta$ meson accounts fairly well
for the mass difference
between negative parity states and the corresponding ground state baryons.
However the predicted masses are always too small by 50--100 MeV, the
difference with the experimental values being larger with SET A parameters.
The qualitative agreement could be partially improved by choosing
a value of $x$ closer to 1, at the price of a worse description of the
hyperfine splitting. The problem, in any case,
cannot be solved within this simple model approximations since they
implicitly imply the constraint $\omega_\eta < m_\eta$ which
is not satisfied phenomenologically
(for instance, $m_{N(1535)}-m_N > m_\eta$).
In order to obtain a better quantitative agreement with the experimental mass
spectrum one has to consider the coupling with the pion vibrational
modes. This ``coupled channels" procedure has already provided
a remarkable improvement
in the predictions of the $\eta$-photoproduction
observables \cite{AHcoup}. It is also clear that this new available
channel will have some influence on the $P$-wave phase shifts.
Whether this is enough to produce some resonance behaviour
in such a channel is however an open question.
Work along these lines is in progress.

\vspace{1.cm}
\noindent
CG thanks {\it Fondazione Della Riccia} for partial supporting during
the completion of this work.

\pagebreak

\pagebreak
{\bf \Large Table and Figure Captions}
\begin{description}

\item [Table 1]: Calculated excitation energies (in MeV)
taken with respect to the nucleon mass as
compared to the experimental ones. SET A corresponds to
$f_\pi = 93 \ {\rm MeV}$, $e = 4.26$ and  $m_\pi = 138 \ {\rm MeV}$ while
SET B corresponds to
$f_\pi = 64.5 \ {\rm MeV}$, $e = 5.45$ and  $m_\pi = 0$. In both cases,
$f_K / f_\pi = 1.22$ and $m_K = 495 \ {\rm MeV}$. For SET A $x$
is taken to the value $x=0.33$ while for SET B to $x=0.66$.

\item [Fig. 1]: The $\eta$ bound state energy as a function of the $x$
parameter. The values of the other parameters are as in Table 1.

\item [Fig. 2]: Eta-soliton phase shifts corresponding to
partial waves with $l \le 3$ as a function of the $\eta$ energy.
The values of the parameters are as in Table 1.

\end{description}

\pagebreak

\begin{center}
{\Large \bf Table 1}
\vspace{1.cm}

\begin{tabular}{|c|c|c|c|c|c|c|c|}
\hline
$n_K$ & $n_\eta$ & $I$ & $J^P$ & SET A & SET B & Emp. & Particle \\ \hline
 0    &    1     & 1/2 & $1/2^-$ & 467   &  517  & 596  & N(1535)  \\
 1    &    0     &  0  & $1/2^+$ & 172   &  165  & 177  & $\Lambda$ \\
 1    &    1     &  0  & $1/2^-$ & 639   &  682  & 731  & $\Lambda(1670)$ \\
 1    &    0     &  1  & $1/2^+$ & 252   &  243  & 254  & $\Sigma$ \\
 1    &    1     &  1  & $1/2^-$ & 719   &  760  & 811  & $\Sigma(1750)$ \\
 2    &    0     & 1/2 & $1/2^+$ & 393   &  377  & 379  & $\Xi$ \\
 2    &    1     & 1/2 & $1/2^-$ & 860   &  894  & 1011 & $\Xi(1950)$ ?\\
\hline
\end{tabular}
\end{center}

\end{document}